\documentstyle[12pt,epsf]{article}
\begin{document}
\begin{flushright}
Fermilab-PUB-96/317-E \\
\end{flushright}
\vspace{0.5in}
\begin{center}
\begin{Large}
{\bf  Measurement of Dijet Angular Distributions at CDF}
\end{Large}
\end{center}
%This gets replaced by authors and institutions
%\vspace{0.5in}
%\begin{center}
%\begin{Large}
%{\bf The CDF Collaboration}
%\vspace{0.5in}
%\end{Large}
%\end{center}
%End of bogus collaboration list
\font\eightit=cmti8
\def\r#1{\ignorespaces $^{#1}$}
\hfilneg
\begin{sloppypar}
\noindent
F.~Abe,\r {15} H.~Akimoto,\r {34}
A.~Akopian,\r {29} M.~G.~Albrow,\r 7 S.~R.~Amendolia,\r {25} 
D.~Amidei,\r {18} J.~Antos,\r {31} C.~Anway-Wiese,\r 4 S.~Aota,\r {34}
G.~Apollinari,\r {29}
% ...
\end{sloppypar}

\vskip .025in
\begin{center}
(CDF Collaboration)
\end{center}

\vskip .025in
\begin{center}
\r 1  {\eightit Argonne National Laboratory, Argonne, Illinois 60439} \\
\r 2  {\eightit Istituto Nazionale di Fisica Nucleare, University of Bologna,
I-40126 Bologna, Italy} \\
\r 3  {\eightit Brandeis University, Waltham, Massachusetts 02254} \\
\r 4  {\eightit University of California at Los Angeles, Los 
Angeles, California  90024} \\  
\r 5  {\eightit University of Chicago, Chicago, Illinois 60637} \\
\r 6  {\eightit Duke University, Durham, North Carolina  27708} \\
\r 7  {\eightit Fermi National Accelerator Laboratory, Batavia, Illinois 
60510} \\
\r 8  {\eightit Laboratori Nazionali di Frascati, Istituto Nazionale di Fisica
               Nucleare, I-00044 Frascati, Italy} \\
\r 9  {\eightit University of Geneva, CH-1211 Geneva 4, Switzerland} \\
\r {10} {\eightit Harvard University, Cambridge, Massachusetts 02138} \\
\r {11} {\eightit Hiroshima University, Higashi-Hiroshima 724, Japan} \\
\r {12} {\eightit University of Illinois, Urbana, Illinois 61801} \\
\r {13} {\eightit Institute of Particle Physics, McGill University, Montreal 
H3A 2T8, and University of Toronto,\\ Toronto M5S 1A7, Canada} \\
\r {14} {\eightit The Johns Hopkins University, Baltimore, Maryland 21218} \\
\r {15} {\eightit National Laboratory for High Energy Physics (KEK), Tsukuba, 
Ibaraki 305, Japan} \\
\r {16} {\eightit Ernest Orlando Lawrence Berkeley National Laboratory, 
Berkeley, California 94720} \\
\r {17} {\eightit Massachusetts Institute of Technology, Cambridge,
Massachusetts  02139} \\   
\r {18} {\eightit University of Michigan, Ann Arbor, Michigan 48109} \\
\r {19} {\eightit Michigan State University, East Lansing, Michigan  48824} \\
\r {20} {\eightit University of New Mexico, Albuquerque, New Mexico 87131} \\
\r {21} {\eightit The Ohio State University, Columbus, OH 43210} \\
\r {22} {\eightit Osaka City University, Osaka 588, Japan} \\
\r {23} {\eightit Universita di Padova, Istituto Nazionale di Fisica 
          Nucleare, Sezione di Padova, I-35131 Padova, Italy} \\
\r {24} {\eightit University of Pennsylvania, Philadelphia, 
        Pennsylvania 19104} \\   
\r {25} {\eightit Istituto Nazionale di Fisica Nucleare, University and Scuola
               Normale Superiore of Pisa, I-56100 Pisa, Italy} \\
\r {26} {\eightit University of Pittsburgh, Pittsburgh, Pennsylvania 15260} \\
\r {27} {\eightit Purdue University, West Lafayette, Indiana 47907} \\
\r {28} {\eightit University of Rochester, Rochester, New York 14627} \\
\r {29} {\eightit Rockefeller University, New York, New York 10021} \\
\r {30} {\eightit Rutgers University, Piscataway, New Jersey 08854} \\
\r {31} {\eightit Academia Sinica, Taipei, Taiwan 11529, Republic of China} \\
\r {32} {\eightit Texas A\&M University, College Station, Texas 77843} \\
\r {33} {\eightit Texas Tech University, Lubbock, Texas 79409} \\
\r {34} {\eightit University of Tsukuba, Tsukuba, Ibaraki 305, Japan} \\
\r {35} {\eightit Tufts University, Medford, Massachusetts 02155} \\
\r {36} {\eightit University of Wisconsin, Madison, Wisconsin 53706} \\
\r {37} {\eightit Yale University, New Haven, Connecticut 06511} \\
\end{center}

\renewcommand{\baselinestretch}{2}
\large
\normalsize
\clearpage

\begin{center}
{\bf Abstract}
\end{center}
We have used $106$ pb$^{-1}$ of data collected in $p\bar{p}$ 
collisions at $\sqrt{s}=1.8$ TeV by the Collider Detector
at Fermilab to measure jet angular distributions in events with two jets
in the final state. The angular 
distributions agree with next to leading order (NLO) predictions of Quantum 
Chromodynamics (QCD) in all dijet invariant mass 
regions. The data exclude at 95\% confidence level (CL) a model of quark 
substructure in which 
only up and down quarks are composite and the contact interaction scale is
$\Lambda^+_{ud}\leq 1.6$ TeV or $\Lambda^-_{ud}\leq 1.4$ TeV. For a model in which all quarks are 
composite the excluded regions are $\Lambda^+\leq 1.8$ TeV and $\Lambda^-\leq 1.6$ TeV.\\
\vspace*{0.3in}
\vspace*{1in}

Hard collisions between protons and antiprotons predominantly produce 
events containing two high energy jets (dijets).
Measurement of the distribution of the scattering angle, between the dijet and
the proton beam in the dijet center of mass frame, can provide a
fundamental test of QCD and a sensitive probe of new physics.
Dijet angular distributions reflect 
the dynamics of the hard scattering of quarks and gluons, and are expected to
be fairly insensitive to the momentum distributions of these partons within 
the proton.  
As in Rutherford scattering, dijet angular distributions from QCD processes are
peaked in the 
forward direction.  In contrast, many sources of new physics produce more
isotropic dijet angular distributions.
Our previous measurements using 4.2 pb$^{-1}$ of data found the dijet angular 
distributions to be in good agreement with QCD 
predictions~\cite{ref_angle_prls}, and 
excluded a compositeness scale $\Lambda^+_{ud}<1.0$ TeV for a contact 
interaction associated with compositeness of up and down 
quarks~\cite{ref_compositeness}. Here 
we report a measurement with a data sample that is 25 times larger.

We are also motivated by the observation
that the inclusive differential jet cross section is above a QCD prediction at 
high transverse energy, $E_T$~\cite{ref_jet_prl_1a}. Interpretations of this 
high-$E_T$ jet excess vary from explanations within the Standard Model 
(modifications of the parton distributions~\cite{ref_cteq2,ref_cteq} or QCD  
corrections~\cite{ref_resum}) to explanations beyond the Standard Model (e.g. 
quark compositeness~\cite{ref_compositeness}, excited quarks~\cite{ref_qstar}, 
new Z bosons~\cite{ref_zprime}, new massive gluons~\cite{ref_coloron}, light
gluinos~\cite{ref_gluino}, and anomalous chromomagnetic moments of 
quarks~\cite{ref_moment}). 
Measurements of the dijet angular 
distributions can help in resolving whether the measured excess of events with
high $E_T$ jets is a signal of new physics or merely new information on the 
ingredients of QCD calculations.

A detailed description of the Collider Detector at Fermilab (CDF) can be found 
elsewhere~\cite{ref_CDF}. We use a coordinate system with $z$ along the proton
beam direction, transverse coordinate perpendicular to the beam, azimuthal angle $\phi$, 
polar angle $\theta$, and pseudorapidity $\eta=-\ln \tan(\theta/2)$. 
Jets are identified as localized energy depositions in the CDF calorimeters, 
which are constructed in a tower geometry.  
 The jet axis is defined as the centroid in ($\eta$,$\phi$)
 space of the calorimeter tower transverse energies 
 inside a radius $R=\sqrt{(\Delta\eta)^2 + (\Delta\phi)^2}=0.7$ of the axis.
The jet energy, $E$, and momentum, $\vec{P}$, are defined as the scalar and 
vector sums, respectively, of the tower energies inside this radius. 
$E$ and $\vec{P}$ are corrected for non-linearities in the calorimeter 
response, energy lost in uninstrumented regions and 
outside the clustering cone, and
energy gained from the underlying event and multiple interactions. 
The jet energy corrections increase the reconstructed jet energies on average 
by 24\%(19\%) for 50 GeV (500 GeV) jets.  Details of jet 
reconstruction and
jet energy corrections can be found elsewhere~\cite{ref_jet}.

The dijet system consists of the 
two jets with the highest transverse momentum in the event (leading jets). 
We measure inclusive dijet events, defined as $p\bar{p} \rightarrow$ 2 leading 
jets + X, where X can be anything, including additional jets.
The dijet invariant mass is defined as 
$M=\sqrt{(E_1 + E_2)^2 - (\vec{P}_1 + \vec{P}_2)^2}$.
We use the dijet angular variable
$\chi=\exp(|\eta_1 - \eta_2|)$, where $\eta_1$ and $\eta_2$ are the
pseudorapidities of the two leading jets.
The variable $\chi$ has the benefit of only containing angular quantities, and
hence is more accurately measured than a variable that involves the absolute 
jet energy. For the case of $2\rightarrow 2$ parton scattering, $\chi$ is
related to the scattering angle in the center of mass frame, $\theta^*$, 
by $\chi=(1+|\cos\theta^*|)/(1-|\cos\theta^*|)$.       
The $\chi$ distribution of QCD produced jets is roughly flat while many models 
of quark
compositeness give angular distributions that are strongly peaked at low 
$\chi$. To select events with high trigger efficiency and to avoid problematic 
regions of the detector, this analysis requires $\chi<5$, $|\eta_1|<2$, and 
$|\eta_2|<2$. To characterize
the shape of the angular distribution in a mass bin with a single number, we
use the variable $R_{\chi} = N(\chi<2.5)/N(2.5<\chi<5)$, 
the ratio of the number of dijet events with $\chi<2.5$ to the number of dijet 
events with $2.5<\chi<5$.  
Isotropic angular distributions and contact interactions 
both tend to produce more events in the region $\chi<2.5$ than QCD, and hence 
will have a higher value of $R_\chi$. The pivot point $\chi=2.5$ 
was chosen to optimize the sensitivity to a left-handed contact 
interaction~\cite{ref_compositeness}.

Our data sample was obtained in the 1992-95 running periods using four 
single-jet 
triggers with thresholds on the uncorrected cluster transverse energies
of 20, 50, 70, and 100 GeV.  After applying the jet energy corrections the last 
three trigger samples were used to measure the dijet angular distribution in 
mass bins above 241, 300, and 400 GeV/c$^2$, respectively.  
At these mass thresholds the trigger efficiencies were 
greater than 95\% at all values of $\chi$ considered; the average efficiency
was greater than 99\% in each mass bin.  The 20 GeV trigger sample was only 
used to measure the trigger efficiency of the 50 GeV sample.
The four data samples 
corresponded to integrated luminosities of $0.126$, $2.84$, $14.1$, and $106$ 
pb$^{-1}$.
To utilize the projective nature of the calorimeter towers, the $z$ position
of the event vertex was required to be within 60 cm of the center of the
detector; this cut removed 7\% of the events. Backgrounds from cosmic rays, 
beam halo, and detector noise were removed with the cuts reported 
previously~\cite{ref_dijet_prl_1a}, and residual backgrounds were removed by 
requiring that the total observed energy be less than 2 TeV. 

The raw $\chi$ distribution was measured in five bins of dijet mass: 
$241<M<300$, $300<M<400$, $400<M<517$, $517<M<625$, and $M>625$ GeV/c$^2$.
Variations in the jet response and energy resolution of the calorimeter as a 
function of detector $\eta$ produced distortions in the measured angular 
distribution. To understand and correct for this effect a parametrized 
Monte Carlo program was developed that  
modeled in detail the measured jet response of the CDF detector after 
the application of the standard jet corrections.  The relative jet response 
was determined from conservation of transverse momentum, $P_T$, by requiring a 
jet in the region $0.15<|\eta|<0.9$ and 
measuring the relative response of a jet in another pseudorapidity region 
(jet $P_T$ balancing).
For jet $P_T$ balancing, events were selected by requiring there be
two and only two jets with $P_T>15$ GeV/c, and that the azimuthal angle
separating the two jets satisfy $150^\circ <\Delta\phi< 210^\circ$.
The largest effect was for $M>625$ GeV/c$^2$, where a 
6\% larger jet response at $|\eta|<0.15$ 
and a 4\% smaller jet response at $0.9<|\eta|<1.4$ produced a tilt in the 
Monte Carlo angular distribution that increased the relative rate at $\chi=1$ 
by about 10\% and lowered the relative rate at $\chi=5$ by about 10\%. We 
corrected both the $\chi$ and $R_\chi$ distributions for these and similar 
effects.  The correction reduced
$R_\chi$ by 1\%, 2\%, 3\%, 5\%, and 6\% for the 5 mass bins, respectively.  
The correction increases with dijet invariant mass because the mass spectrum 
is steeper at higher mass values, leading to a larger distortion of the 
angular distribution.

\begin{figure}[tbh]
\hspace*{-0.5in}
\vspace*{-1.0in}
\epsffile[36 81 540 650]{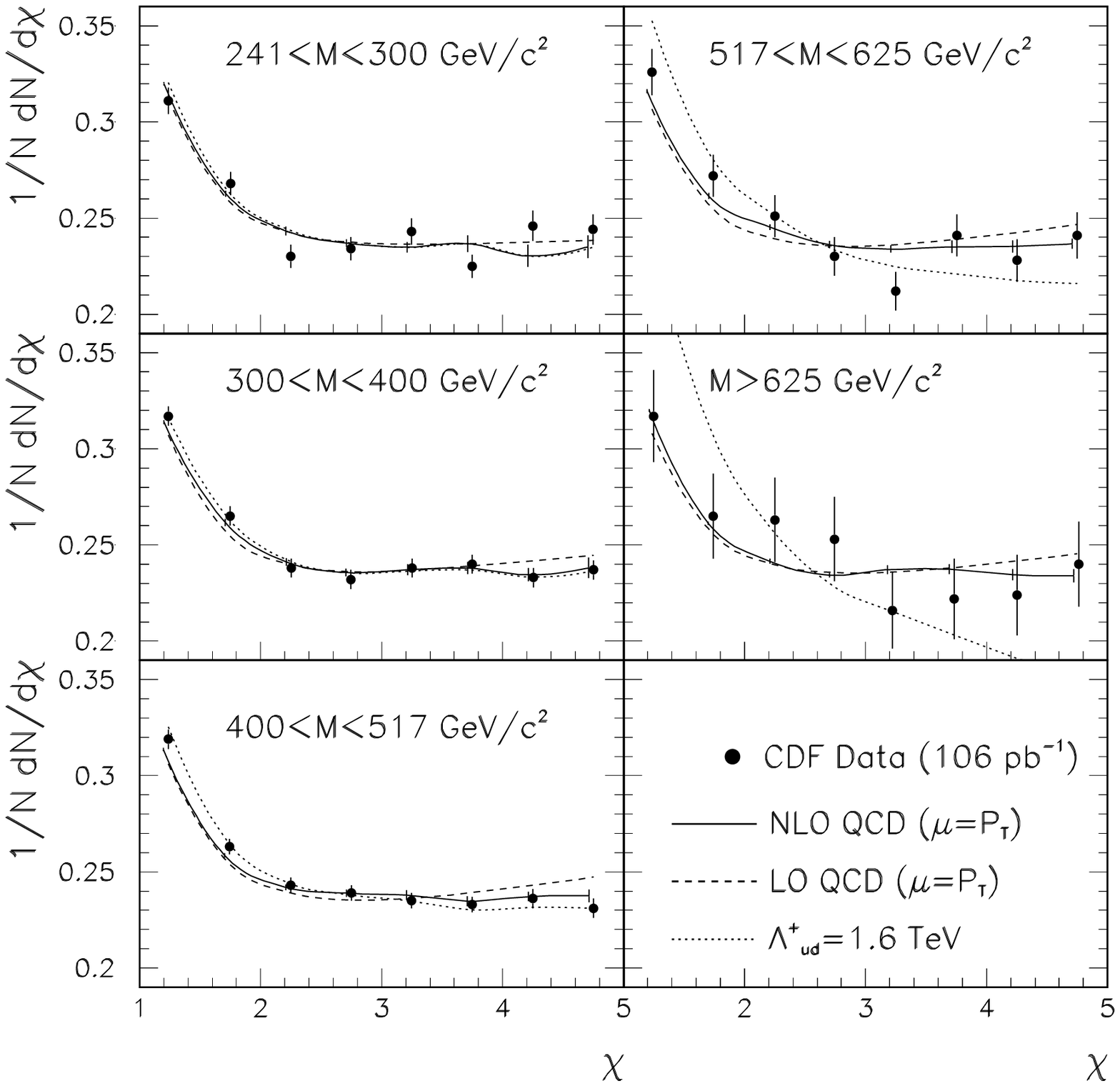}
\caption{ 
The dijet angular distribution (points) compared to predictions of NLO QCD 
(solid curve), LO QCD (dashed curve), and LO QCD with a quark contact 
interaction (dotted curve).
The contact interaction calculation is normalized to equal NLO QCD when 
$\Lambda_{ud}=\infty$ (see text). Error bars on the data and NLO QCD are 
statistical.}
\label{fig_chi}
\end{figure}

\begin{figure}[tbh]
\hspace*{-0.3in}
\vspace*{-1.0in}
\epsfysize=7.0in
\epsffile[36 81 540 650]{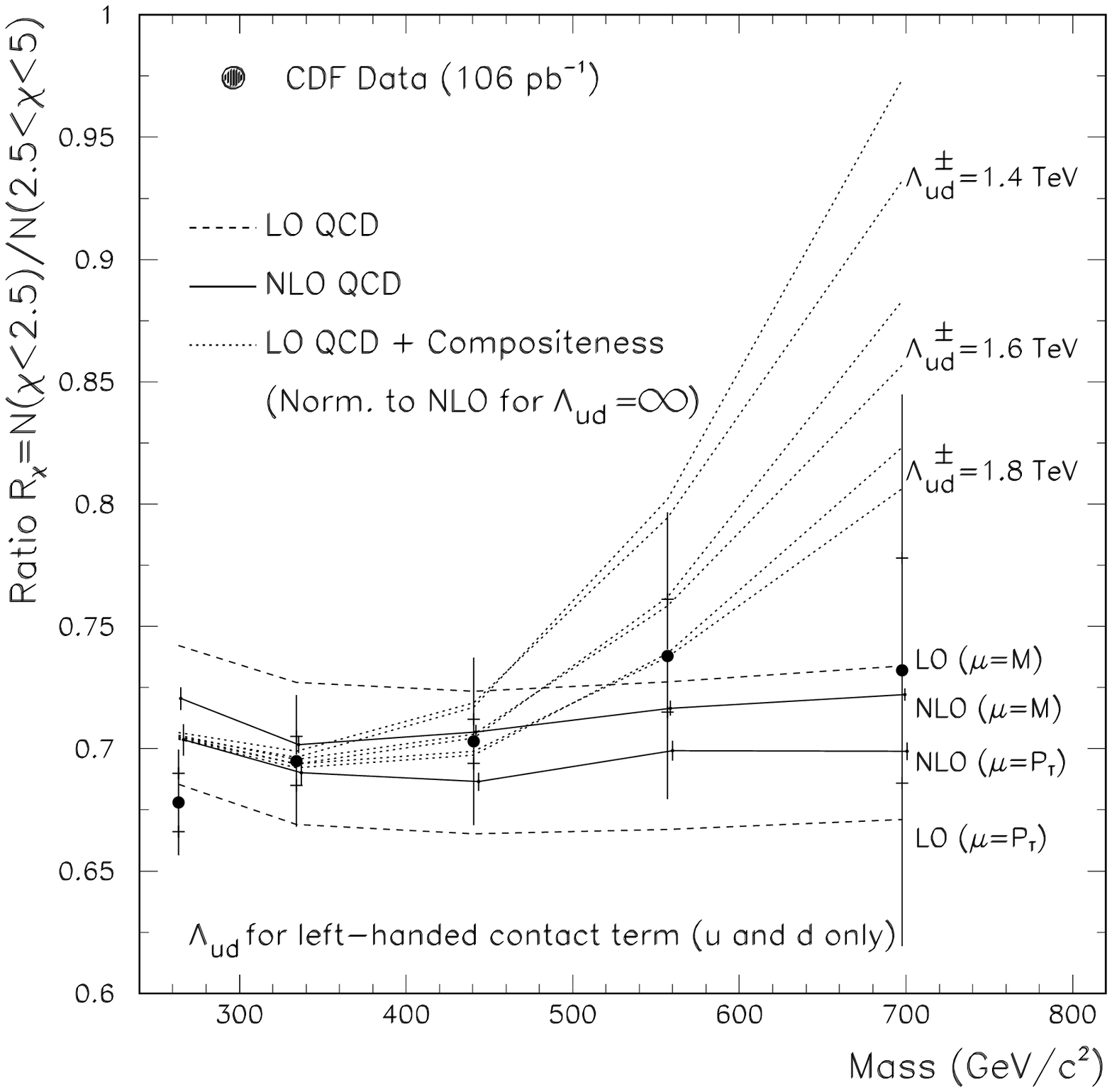}
\caption{The dijet angular ratio (points) as a function of the dijet
invariant mass, compared to LO QCD (dashed curve), 
NLO QCD (solid curve), and LO QCD with a quark contact interaction normalized 
to NLO at $\Lambda_{ud}=\infty$ (dotted curve). QCD is shown for two 
renormalization scales
($\mu=M$ and $\mu=P_T$). Contact interactions are displayed for three
different compositeness scales, with two different signs for the amplitude
of the contact term (upper dotted curve is $\Lambda^+_{ud}$, lower dotted curve
is $\Lambda^-_{ud}$). The inner error bars on the data are statistical uncertainties
and the outer error bars are statistical and systematic uncertainties added
in quadrature. The error bars on NLO QCD are statistical.}
\label{fig_ratio}
\end{figure}

In Figs.~\ref{fig_chi} and \ref{fig_ratio} and Tables I and II we present the 
corrected $\chi$ and $R_\chi$ distributions. The data are 
compared to the parton level predictions of leading order (LO) QCD, next to
leading order (NLO) QCD from the JETRAD Monte Carlo program~\cite{ref_jetrad}, 
and QCD plus a contact
interaction. For the contact interaction curve, 
we normalized LO QCD plus a contact interaction to equal the NLO QCD prediction
with renormalization scale $\mu=P_T$ when the contact scale is 
$\Lambda=\infty$. 
This was done by multiplying the prediction, from LO QCD plus 
a contact interaction, by the ratio of the NLO to LO QCD predictions.
The LO calculations use CTEQ2L 
parton distributions, and the NLO QCD calculation uses CTEQ2M parton 
distributions~\cite{ref_cteq_pdf}. Alternate parton distribution sets 
were tried, 
including one in which the gluon distribution of the proton was
significantly increased~\cite{ref_cteq2}, and the 
calculations were insensitive to the choice of the parton 
distribution. In Fig.~\ref{fig_ratio} the QCD calculations are shown for two 
different choices of 
renormalization scale, $\mu=M$ and $\mu=P_T$. The vertical axis in
Fig.~\ref{fig_ratio}, $R_\chi$, describes the shape of the angular distribution 
at a fixed mass, and is sensitive to the renormalization scale choice.
The choice $\mu=M$
makes $\mu$ constant as a function of $\chi$ in a bin of fixed mass, while the 
choice $\mu=P_T$ requires $\mu$ to vary with $\chi$. 
The renormalization scale dependence of the
NLO calculation, which is significantly less than that of the LO calculation,
provides an estimate of the uncertainty in the NLO QCD calculation.
Figure~\ref{fig_ratio} also illustrates that a contact interaction would 
cause $R_\chi$ to increase at high mass, while QCD calculations predict that 
$R_\chi$ is roughly 0.7 at all masses shown.  
The angular distributions and angular ratio are in good 
agreement with the NLO QCD prediction.

The systematic uncertainties, shown only in Fig.~\ref{fig_ratio} and Table II, 
arise from the uncertainty in the jet energy response as a function of $\eta$. 
The response uncertainties are largest in the region $|\eta|<0.15$
(between 3\% and 6\%) and the region $0.9<|\eta|<1.4$ (4\%).
Other systematic uncertainties are negligible in comparison.  Since the 
systematic uncertainties are larger than the statistical uncertainties, a 
cross check was performed to verify the integrity of the measurement. The
uncorrected $\chi$ and $R_\chi$ distributions were remeasured with the detector
pseudorapidity requirement $0.1<|\eta|<1.0$, eliminating the most problematic
regions of the detector.  The resulting uncorrected distributions were then 
corrected back to the standard region $|\eta|<2$, using the same parametrized 
Monte Carlo program, and
compared with the standard results.  The corrected $\chi$ and $R_\chi$ 
distributions for the region $0.1<|\eta|<1.0$ agreed with those determined in 
the standard region $|\eta|<2$ within the statistical uncertainties. 

The systematic uncertainties on $R_\chi$ are highly correlated as a function of
mass.
The diagonal terms of the covariance matrix for the $R_\chi$ vs. mass 
distribution can be written as $V_{ii} = \sigma^2_i(stat) + 
\sigma^2_i(sys)$, and the off-diagonal terms are 
$V_{ij} = \sigma_i(sys) \sigma_j(sys)$, $i\neq j$, for mass bins $i$ and $j$.
Using this prescription and Table II the reader can reconstruct the full 
covariance matrix.
We form a statistical comparison between the data and the theory by using 
the inverse of the covariance matrix, $(V^{-1})_{ij}$, and the difference 
between the data and the theory in each bin, $\Delta_i$, to define 
$\chi^2={\displaystyle \sum_{i,j}} \Delta_i (V^{-1})_{ij}
\Delta_j$.  The resulting comparison between data and NLO QCD 
with renormalization scale $\mu=P_T$ is $\chi^2=8.36$ for 5 degrees of freedom.
This is a better agreement than $\chi^2=13.1$ for NLO QCD with $\mu=M$. 

We exclude at 95\% CL any theoretical prediction which gives a $\chi^2$ of 
greater than 11.1 when compared to our data. In a model of contact 
interactions where the
up and down type quarks are composite we exclude
at 95\% CL the scales $\Lambda^+_{ud}\leq 1.6$ TeV and $\Lambda^-_{ud}\leq1.4$ 
TeV.  For flavor symmetric contact 
interactions among all quark flavors~\cite{ref_lane}, not just up and down 
quarks, the scales excluded by the angular distribution are 
$\Lambda^+\leq1.8$ TeV and $\Lambda^-\leq1.6$ TeV. 

We compare these exclusions with the inclusive jet cross section 
analysis~\cite{ref_jet_prl_1a}, where we 
reported a broad minimum in the $\chi^2$ between data and the compositeness 
model for the scales  $1.5 \leq \Lambda^+_{ud}\leq 1.8$ TeV, and best agreement 
for the scale $\Lambda^+_{ud} = 1.6$ TeV. 
Here we have excluded at 95\% CL the 
portion of this broad minimum up to the scale $\Lambda^+_{ud} = 1.6$ TeV.
The inclusive jet cross section $\chi^2$ is sensitive to the choice of parton 
distributions while the dijet angular distribution $\chi^2$ is not.  The 
inclusive jet analysis used MRSD0$^\prime$ parton distributions~\cite{ref_mrs} 
with renormalization scale $\mu = E_T/2$.
Changing the sign of the contact interaction from positive to negative 
produces a larger deviation of the composite model from QCD for the inclusive 
jet cross section, but produces a smaller deviation from QCD in the shape of 
the angular distribution.
Therefore, if we had chosen to fit the inclusive jet cross section with the
negative sign contact interaction, best agreement would have been found with 
less compositeness, corresponding to the larger scale $\Lambda^-_{ud}=1.8$ TeV, 
which is not excluded by the angular distribution. 

In conclusion, we have measured the dijet angular distributions and found them
to be in good agreement with NLO QCD. We have presented limits on the 
left-handed contact interactions among quarks that could result if quarks were 
composite particles. 
Although the origin of the excess in the inclusive jet $E_T$ spectrum has 
not been determined, the angular distribution data are consistent with 
the hypothesis that the high-$E_T$ jet excess is caused by effects within 
the Standard Model. The angular distribution data exclude at the 95\% CL 
the hypothesis that the high-$E_T$ jet excess is caused by a 
contact interaction among up and down quarks with scale $\Lambda^+_{ud} 
\leq 1.6$ TeV.                              

    We thank the Fermilab staff and the technical staffs of the
participating institutions for their vital contributions.  This work was
supported by the U.S. Department of Energy and National Science Foundation;
the Italian Istituto Nazionale di Fisica Nucleare; the Ministry of Education,
Science and Culture of Japan; the Natural Sciences and Engineering Research
Council of Canada; the National Science Council of the Republic of China;
and the A. P. Sloan Foundation.

\clearpage

Table I: The dijet angular distribution and statistical uncertainty for the 
five mass bins (GeV/c$^2$).\\

\renewcommand{\baselinestretch}{1.4}
\large
\normalsize

\begin{table}[h]
\begin{center}
\begin{tabular}{|c|c|c|c|c|c|}\hline 
 &  \multicolumn{5}{c|}{(100/N)(dN/d$\chi$)} \\
 $\chi$ & $241<M<300$ & $300<M<400$ & $400<M<517$ &   $517<M<625$ & $M>625$\\ \hline
1.25& $31.1\pm0.7$ & $31.7\pm0.5$ & $31.9\pm0.5$ & $32.6\pm1.2$ & $31.7\pm2.4$ \\
1.75& $26.8\pm0.6$ & $26.5\pm0.5$ & $26.3\pm0.4$ & $27.2\pm1.1$ & $26.5\pm2.2$ \\ 
2.25& $23.0\pm0.6$ & $23.8\pm0.5$ & $24.3\pm0.4$ & $25.1\pm1.1$ & $26.3\pm2.2$ \\ 
2.75& $23.4\pm0.6$ & $23.2\pm0.5$ & $23.9\pm0.4$ & $23.0\pm1.0$ & $25.3\pm2.2$ \\ 
3.25& $24.3\pm0.7$ & $23.8\pm0.5$ & $23.5\pm0.4$ & $21.2\pm1.0$ & $21.6\pm2.0$ \\ 
3.75& $22.5\pm0.6$ & $24.0\pm0.5$ & $23.3\pm0.4$ & $24.1\pm1.1$ & $22.2\pm2.1$ \\ 
4.25& $24.6\pm0.8$ & $23.3\pm0.5$ & $23.6\pm0.5$ & $22.8\pm1.1$ & $22.4\pm2.1$ \\ 
4.75& $24.4\pm0.8$ & $23.7\pm0.5$ & $23.1\pm0.5$ & $24.1\pm1.2$ & $24.0\pm2.2$ \\ 
\hline
\end{tabular}
\end{center}
\end{table}

\renewcommand{\baselinestretch}{2}
\large
\normalsize
\vspace*{0.5in}
Table II: The mean dijet mass (GeV/c$^2$), number of events, dijet angular
ratio $R_{\chi}$, and its statistical and systematic uncertainty. The
completely correlated systematic uncertainty can be used to form the
covariance matrix (see text).

\renewcommand{\baselinestretch}{1.4}
\large
\normalsize

\begin{table}[h]
\begin{center}
\begin{tabular}{|c|r|c|c|c|}\hline
$<$Mass$>$ & Events & $R_{\chi}$ & Stat. &  Sys. \\
\hline
263 & 15023 & 0.678 &  0.012 &  0.018\\
334 & 23227 & 0.695 &  0.010 &  0.025\\
440 & 28202 & 0.703 &  0.009 &  0.033\\
557 &  4425 & 0.738 &  0.023 &  0.054\\
698 &  1056 & 0.732 &  0.046 &  0.103\\
\hline
\end{tabular}
\end{center}
\end{table}

\clearpage

\clearpage

\end{document}